%% file: 0_CQA.tex
\documentclass[sigconf]{acmart}
 \usepackage{textcomp, libertine}
\usepackage{booktabs} 
\usepackage[normalem]{ulem}
\usepackage{multirow}
\usepackage{bm}
\useunder{\uline}{\ul}{}
\usepackage{subcaption}
\usepackage[para]{footmisc}
\usepackage{enumitem}
\usepackage[ruled]{algorithm2e}
\usepackage{amssymb}
\usepackage{hyperref}
\hypersetup{
,bookmarksnumbered=true%
,hypertexnames=false%
,breaklinks=true%
,colorlinks=true%
,linkcolor=black%
,citecolor=black%
,urlcolor=black%
}

\definecolor{tealblue}{rgb}{0.21, 0.46, 0.53}
\definecolor{wildstrawberry}{rgb}{1.0, 0.26, 0.64}
\definecolor{ao(english)}{rgb}{0.0, 0.5, 0.0}

\def\ignore#1{}

\begin{document}

\copyrightyear{2019} 
\acmYear{2019} 
\setcopyright{acmcopyright}
\acmConference[SIGIR '19]{Proceedings of the 42nd International ACM SIGIR Conference on Research and Development in Information Retrieval}{July 21--25, 2019}{Paris, France}
\acmBooktitle{Proceedings of the 42nd International ACM SIGIR Conference on Research and Development in Information Retrieval (SIGIR '19), July 21--25, 2019, Paris, France}
\acmPrice{15.00}
\acmDOI{10.1145/3331184.3331341}
\acmISBN{978-1-4503-6172-9/19/07}

\settopmatter{printacmref=false, printfolios=false}
\fancyhead{}

\title{BERT with History Answer Embedding for Conversational Question Answering}







\author{Chen Qu$^1$ \quad  Liu Yang$^1$ \quad Minghui Qiu$^2$ \quad  W. Bruce Croft$^1$ \quad Yongfeng Zhang$^3$ \quad Mohit Iyyer$^1$}

\affiliation{%
	\institution{
		$^1$ University of Massachusetts Amherst \quad
		$^2$ Alibaba Group \quad
        $^3$ Rutgers University} 
}
\email{{chenqu, lyang, croft, miyyer}@cs.umass.edu, minghui.qmh@alibaba-inc.com, yongfeng.zhang@rutgers.edu}

\begin{abstract}

Conversational search is an emerging topic in the information retrieval community. One of the major challenges to multi-turn conversational search is to model the conversation history to answer the current question. Existing methods either prepend history turns to the current question or use complicated attention mechanisms to model the history. We propose a conceptually simple yet highly effective approach referred to as \textit{history answer embedding}. It enables seamless integration of conversation history into a conversational question answering (ConvQA) model built on BERT (Bidirectional Encoder Representations from Transformers). We first explain our view that ConvQA is a simplified but concrete setting of conversational search, and then we provide a general framework to solve ConvQA. We further demonstrate the effectiveness of our approach under this framework. Finally, we analyze the impact of different numbers of history turns under different settings to provide new insights into conversation history modeling in ConvQA.


\end{abstract}


\maketitle

\input{1_introduction}
\input{2_related_work}
\input{3_our_approach}

\input{4_experiments}
\input{5_conclusions}

\begin{acks}
This work was supported in part by the Center for Intelligent Information Retrieval and in part by NSF IIS-1715095. Any opinions, findings and conclusions or recommendations expressed in this material are those of the authors and do not necessarily reflect those of the sponsor.
\end{acks}

\bibliographystyle{ACM-Reference-Format}
\bibliography{acmart} 

\end{document}

%% file: 1_introduction.tex
{\fontsize{8pt}{8pt} \selectfont
\textbf{ACM Reference Format:}\\
Chen Qu, Liu Yang, Minghui Qiu, W. Bruce Croft, Yongfeng Zhang, and Mohit Iyyer. 2019. BERT with History Answer Embedding for Conversational Question Answering. In \textit{Proceedings of the 42nd Int'l ACM SIGIR Conference on Research and Development in Information Retrieval (SIGIR'19), July 21--25, 2019, Paris, France.} ACM, NY, NY, USA, 4 pages. \url{https://doi.org/10.1145/3331184.3331341}}
\section{Introduction}
\label{sec:intro}

A long-term goal of the information retrieval (IR) community has been to design a search system that can retrieve information iteratively and interactively~\cite{Belkin1994CasesS,i3r,Kotov2010TowardsNQ}. The emerging field of conversational AI has impacted this goal, leading to a direction referred to as conversational search.
Conversational AI consists of three branches, namely, task-oriented bots, social bots, and question answering (QA) bots~\cite{Gao2018NeuralAT}. The first two have attracted extensive research efforts in the recent years, 
resulting in a wide range of personal assistants, such as Siri and Cortana.
These systems, however, are not capable of handling complicated information-seeking conversations that require multiple turns of information exchange. Much work remains to empower common users to conduct conversational search.

It is natural for people to seek information through conversations. In the setting of conversational search, a user initiates a conversation with a specific information need. The search system conducts multiple turns of interaction with the user via a ``System Ask, User Respond'' paradigm~\cite{Zhang2018TowardsCS} to better understand this information need.
The system then tries to fulfill this need by retrieving answers iteratively based on the user's feedback or clarifying questions. 
The user sometimes asks follow-up questions with a related but new information need and thus enters the next ``cycle'' of the conversational search process. In order to understand the user's latest information need, the system should be capable to handle the conversation history.
In our view, conversational question answering (ConvQA) is a simplified setting of conversational search since ConvQA systems do not focus on asking proactively. However, ConvQA is concrete enough for IR researchers to work on modeling the change of information needs between cycles.
Therefore, we focus on handling conversation history in a ConvQA setting.

ConvQA can be formalized as the relatively well-studied machine comprehension (MC) problem~\cite{quac,coqa,sdnet}. This is achieved by incorporating the conversation history into an MC model. There are two aspects to handle this. The first is history selection, which selects a subset of the history turns that are more helpful than others. The second is history modeling, which models the selected history turns in an MC model.
Thus, we define a general framework to describe these two aspects and lay the groundwork for future efforts with ConvQA. We focus on the history modeling aspect in this work and adopt a rule-based method for history selection.

History modeling is essential for ConvQA. Table~\ref{tab:example} shows a part of a dialog from a ConvQA dataset (QuAC~\cite{quac}). When the user issues the query $\text{Q}_2$, we expect the agent to refer to $\text{A}_1$ so that it can understand the meaning of ``that way''.
In such cases, previous history turns play an essential role in understanding the user's current information need. 

\begin{table}[htbp]
\caption{A part of an information-seeking dialog from QuAC. ``R'', ``U'', and ``A'' denotes role, user, and agent respectively.
}
\label{tab:example}
\footnotesize
\tabcolsep=0.08cm
\vspace{-0.45cm}
\begin{tabular}{@{}clcl@{}}
\toprule
\multicolumn{4}{l}{Topic: Augusto Pinochet: Intellectual life}                                                                                                             \\ \midrule
\multicolumn{1}{c|}{\#}                 & \multicolumn{1}{l|}{ID}  & \multicolumn{1}{c|}{R} & \multicolumn{1}{c}{Utterance}                                            \\ \midrule
\multicolumn{1}{c|}{\multirow{2}{*}{1}} & \multicolumn{1}{l|}{$\text{Q}_1$}  & \multicolumn{1}{c|}{U}     & Was he known for being intelligent                                       \\
\multicolumn{1}{c|}{}                   & \multicolumn{1}{l|}{$\text{A}_1$}  & \multicolumn{1}{c|}{A}     & No, Pinochet was publicly known as a man with a lack of culture.         \\ \midrule
\multicolumn{1}{c|}{\multirow{2}{*}{2}} & \multicolumn{1}{l|}{$\text{Q}_2$}  & \multicolumn{1}{c|}{U}     & Why did people feel that way?                                            \\
\multicolumn{1}{c|}{}                   & \multicolumn{1}{l|}{$\text{A}_2$}  & \multicolumn{1}{c|}{A}     & reinforced by the fact that he also portrayed himself as a common man    \\ \bottomrule
\end{tabular}
\end{table}

Some existing methods simply prepend history turns~\cite{coqa,sdnet} or mark answers in the passage~\cite{quac}. These methods cannot handle a long conversation history. Another existing method~\cite{flowqa} uses complicated attention mechanisms to model history and thus generates relatively large system overhead.
We propose a \textit{history answer embedding} method to model conversation history. Our method is conceptually simple, robust, effective, and has better training efficiency compared to previous approaches.
Moreover, our method is specifically tailored for BERT-based architectures to leverage this latest breakthrough in large scale pre-trained language modeling.

We summarize our contribution as follows. 
(1) We introduce a general framework to handle the conversation history in ConvQA, laying the groundwork for future efforts with this task.
(2) Our proposed history modeling method is one of the first attempts to model conversation history in a BERT-based model for information-seeking conversations.\footnote{Most existing models are tested on CoQA, which is not information-seeking (see Section~\ref{sec:relatedwork}). Moreover, descriptions of these models are not available at the time of our paper submission.} We conduct extensive experiments on QuAC, a large open benchmark, to show the effectiveness of our method. Our methods achieved an F1 score of 62.4 on the QuAC leaderboard\footnote{http://quac.ai/} with a significantly shorter training time compared with the state-of-the-art method. Our code is open sourced.\footnote{\url{https://github.com/prdwb/bert_hae}}
(3) We perform an in-depth analysis to show the impact of different amounts of conversation history. We show that history prepending methods degrade dramatically with long history while our method is robust and shows advantages under such a situation, which provides new insights into conversation history modeling in ConvQA.



%% file: 2_related_work.tex
\section{Related Work}
\label{sec:relatedwork}


ConvQA is closely related to machine comprehension.
High quality datasets~\cite{squad,Marco} have boosted research progress, resulting in a wide range of MC models~\cite{bidaf,drqa}. 
A major difference between ConvQA and MC is that questions in ConvQA are organized in conversations. Thus, we need to model conversation history to understand the current question.
Compared to existing methods that prepend history turns~\cite{coqa,sdnet} to the current question or mark history answers in the passage~\cite{quac}, our method can handle longer conversation history and thus is more robust and effective. In addition, our method is conceptually simple and more efficient than FlowQA~\cite{flowqa} that uses complicated recurrent structures.
Our method is specifically tailored for BERT~\cite{bert}, which pre-trains language representations with bidirectional encoder representations from transformers~\cite{transformer}.

CoQA~\cite{coqa} and QuAC~\cite{quac} are ConvQA datasets with very different properties. Questions in CoQA are often factoid with simple entity-based answers while QuAC consists of mostly non-factoid QAs. More importantly, information-seekers in QuAC have access to the title of the passage only, simulating an information need. The information-seeking setting in QuAC is more in line with our interest as IR researchers. Thus, we focus on QuAC in this work.

In addition to ConvQA, there are other related works focused on conversational search. For example, neural approaches are widely adopted to train a model to ask questions proactively~\cite{Zhang2018TowardsCS}, predict user intent~\cite{UserIntentPred}, predict next question~\cite{Yang2017NeuralMM}, and incorporate external knowledge in response ranking~\cite{Yang2018ResponseRW}. In addition, several observational studies are also conducted~\cite{Qu2018AnalyzingAC,Trippas2018InformingTD} to inform the design of conversational search systems. We focus on dealing with conversation history in this work, which is an integral part in the joint effort of building functional conversational search systems.

%% file: 3_our_approach.tex
\section{Our Approach}
\label{sec:our-approach}
\subsection{Task Definition}
\label{subsec:task}
The ConvQA task is defined as follows~\cite{quac,coqa}. Given a passage $p$, the $k$-th question $q_{k}$ in a dialog, and the conversation history $\mathbf{H}_k$ preceding $q_k$, the task is to answer $q_{k}$ by predicting an answer span $a_k$ within $p$. $\mathbf{H}_k$ has $k-1$ turns, where the $i$-th turn is $\mathbf{H}_k^i = (q_i, a_i)$.
$q_i$ and $a_i$ denote the question and the ground truth answer.

\subsection{A ConvQA Framework}
\label{subsubsec:unified-view}

\begin{figure}
    \centering
    \includegraphics[width=0.45\textwidth]{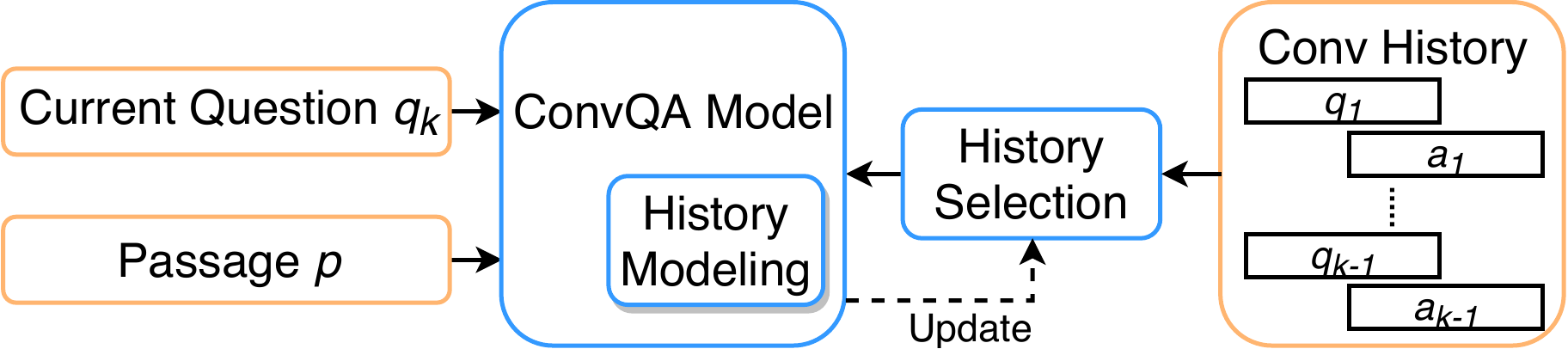}
    \vspace{-0.42cm}
    \caption{A general framework for ConvQA. Orange denotes model input and blue denotes model components.
    }
    \label{fig:framework}
\end{figure}

We present an abstract framework for ConvQA with modularized design in Figure~\ref{fig:framework}. It consists of three major components, a ConvQA model, a history selection module, and a history modeling module. In practice, the history modeling module can be a mechanism inside the ConvQA model. Given a training instance $(p, q_k, \mathbf{H}_k, a_k)$, the history selection module chooses a subset of the history turns $\mathbf{H}_k'$ that are expected to be more helpful than others. The history modeling module then incorporates  $\mathbf{H}_k'$ into the ConvQA model. If the history selection module is a learned policy, the ConvQA model can generate a signal to guide its update. 
In this work, we employ a simple rule as the history selection module that always chooses the immediate $j$ previous turn(s). This is based on the intuition that closer history turns are typically more relevant to the current question. 
We introduce our implementations for the ConvQA model and the history modeling module in the following sections.

\subsection{BERT with History Answer Embedding}
\label{subsubsec:cqa-model}
Our implementation for the ConvQA model can be considered as an MC model integrated with a history modeling mechanism.

\subsubsection{\textbf{Machine Comprehension}}
\label{subsubsec:machine-comprehension}
Our model is adapted from the BERT-based MC model by \citet{bert}. The input is a question and a passage, and the output is the probability of passage tokens being the start/end token of the answer span.
We illustrate the model architecture in Figure~\ref{fig:cqa-model}. First, the question and the passage are packed into a sequence. Then BERT generates a representation for each token based on the embeddings for tokens, segments, and positions. After that, a start/end vector is learned to compute the probability of a token being the start/end token of the answer span. Specifically, let $\mathbf{T}_i$ be the BERT representation of the $i$-th token and $\mathbf{S} \in \mathbb{R}^h$ be the start vector, where $h$ is the token representation size. The probability of this token being the start token is $P_i = \frac{e^{\mathbf{S} \cdot \mathbf{T}_i}}{\sum_j e^{\mathbf{S} \cdot \mathbf{T}_j}}$. 
The probability of a token being the end token is computed likewise. The loss is the average of the cross entropy loss for the start and end positions. 
Invalid predictions are discarded at testing time.


\subsubsection{\textbf{History Answer Embedding}}
\label{subsubsec:cqa-in-context}
\begin{figure}
    \centering
        \includegraphics[width=0.45\textwidth]{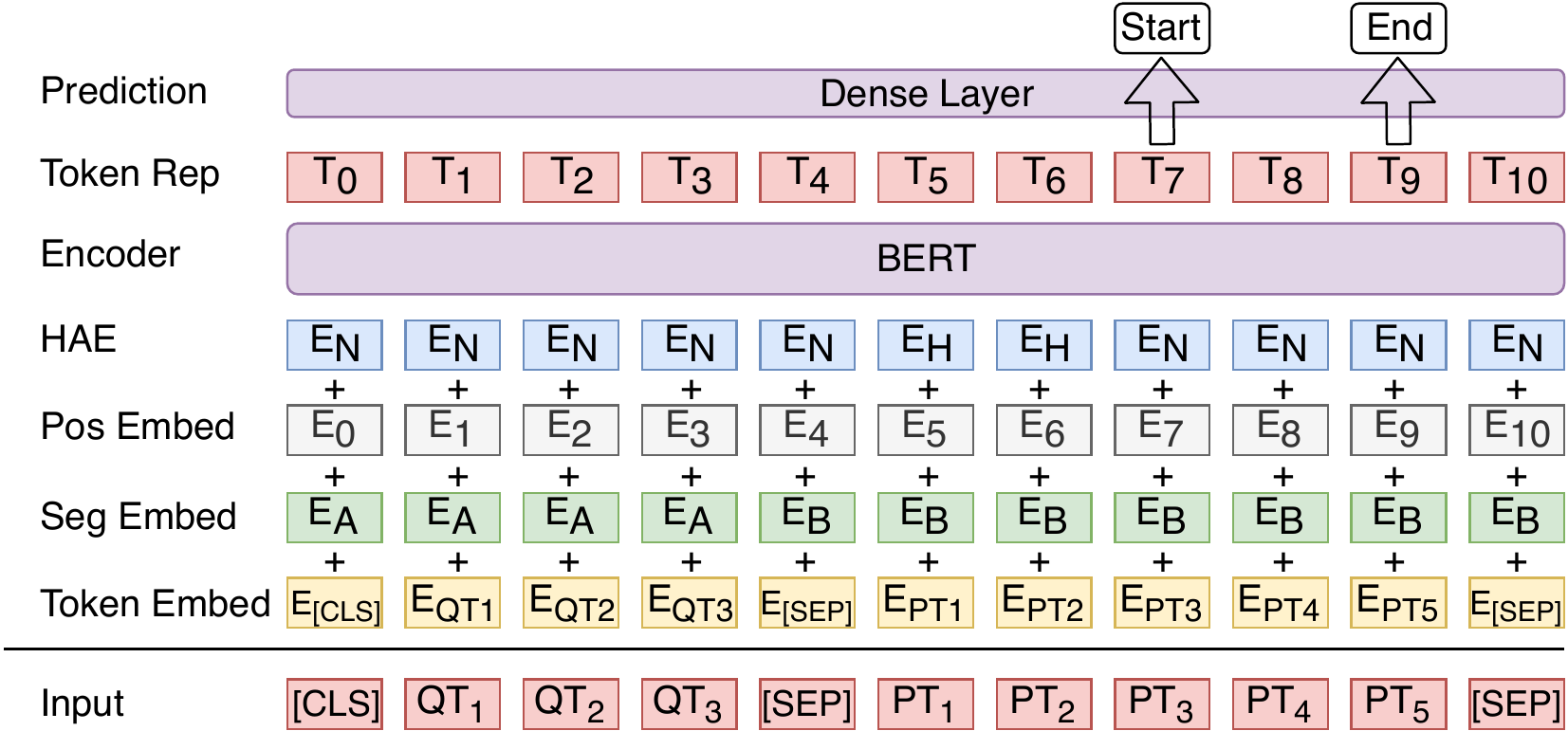}
    \vspace{-0.4cm}
    \caption{Architecture of the ConvQA model with HAE. $E_H$/$E_N$ in HAE denote the token is in/not in history answers.
    }
    \label{fig:cqa-model}
\end{figure}

One important difference of MC and ConvQA lies in handling conversation history.
Suppose we are given a subset of the conversation history chosen by the history selection module for the current question. There are various ways to model the selected history turns. The most intuitive way is to prepend the conversation history to the current question~\cite{coqa,sdnet}.
In this work, we propose a different approach to model the conversation history by giving tokens extra embedding information.
As shown in Figure~\ref{fig:cqa-model}, a \textit{history answer embedding} (HAE) layer is included in addition to other embeddings. 
We learn two unique history answer embeddings that denote whether a token is part of history answers or not.
This introduces the conversation history to BERT in a natural way. HAE modifies the embedding information for a token and thus has influence on the token representation generated by BERT, not only for this token but also for other tokens since BERT considers contextual information. This process also improves the prediction of the answer span as shown in the experiments. By representing conversation history with HAE, we turn an MC model into a ConvQA model.

\subsubsection{\textbf{Model Training}}
\label{subsubsec:cqa-training}

Given a training instance $(p, q_k, \mathbf{H}_k, a_k)$, we first transform it to a list of variations, where each variation $(p, q_k, \mathbf{H}_k^i, a_k)$ contains only one turn of the conversation history.
A history selection module then selects immediate previous $j$ turns. 
After that, we merge the selected variations to form a new instance $(p, q_k, \mathbf{H}_k', a_k)$. It is then used to generate input for the ConvQA model, where $\mathbf{H}_k'$ is used for HAE.
We use a sliding window approach to split long passages following \citet{bert}.


%% file: 4_experiments.tex
\section{Experiments}
\label{sec:exp}


\subsection{Data Description}
\label{subsec:data}
We experiment with QuAC (Question Answering in Context)~\cite{quac}. 
This dataset contains interactive dialogs between information-seekers and -providers. The seeker tries to learn about a \textit{hidden} Wikipedia passage by asking questions. He/she has access to the heading of the passage only. The provider answers the questions by giving a short span of the passage.
Many questions have co-references with conversation history.
The training/validation sets have over 11K/1K dialogs with 83K/7K questions. 
All dialogs are within 12 turns, meaning that a question can have at most 11 history turns. 



\subsection{Experimental Setup}
\label{subsec:setup}
\subsubsection{\textbf{Competing Methods}}
\label{subsubsec:baselines}
We consider all the methods on the QuAC leaderboard as baselines. The competing methods are:
\begin{itemize}[leftmargin=*, noitemsep, topsep=0pt]
    \item \textbf{BiDAF++}~\cite{quac}: BiDAF++ augments BiDAF~\cite{bidaf} with self-attention and contextualized embeddings.
    
    \item \textbf{BiDAF++ w/ 2-Context}~\cite{quac}: It incorporates 2 history turns in BiDAF++ by encoding the dialog turn \# in question embeddings and concatenating marker embeddings to passage embeddings. 
    
    \item \textbf{FlowQA}~\cite{flowqa}: It considers conversation history by integrating intermediate representation generated when answering previous questions and thus can grasp the latent semantics of the history.
    
    \item \textbf{BERT}: We implement a ConvQA model with BERT as described in Section~\ref{subsubsec:machine-comprehension}. This version is without any history modeling.
    
    \item \textbf{BERT + Prepend History Turns}: On top of BERT, we consider conversation history by prepending history turn(s) to the current question. \textbf{BERT + PHQA} prepends both history questions and answers; \textbf{BERT + PHA} prepends history answers only.
    
    \item \textbf{BERT + History Answer Embedding (HAE)}: A BERT-based ConvQA model with our history answer embedding method. 
\end{itemize}

\subsubsection{\textbf{Evaluation Metrics}}
\label{subsubsec:metrics}

We use the word-level F1 to evaluate the overlap of the prediction and the ground truth answer and HEQ (human equivalence score) to measure the percentage of examples where system F1 exceeds/matches human F1. HEQ is computed on a question level (HEQ-Q) and a dialog level (HEQ-D). 

\subsubsection{\textbf{Implementation Details}}
\label{subsubsec:details}
Models are implemented with TensorFlow.\footnote{\url{https://www.tensorflow.org/}} We use the BERT-Base (Uncased) model\footnote{\url{http://goo.gl/language/bert}} with the max sequence length set to 384. The batch size is set to 12. The number of history turns to incorporate is tuned as presented in Section~\ref{subsec:impact-of-history}. We train the ConvQA model with an Adam weight decay optimizer with an initial learning rate of 3e-5. We set the stride in the sliding window for passages to 128, the max question length to 64, and the max answer length to 30. We save checkpoints every 1,000 steps and test on the validation set. We use QuAC v0.2.

\subsection{Main Evaluation Results}
\label{subsec:results}
Experiment results are shown in Table~\ref{tab:results}. Our best model was evaluated officially and the result is displayed on the leaderboard\footnote{\url{http://quac.ai/}}.

\begin{table}[htbp]
\caption{Evaluation results. Each cell displays val/test scores.
Test results are from the QuAC leaderboard on 02/17/2019. $\ddagger$ means statistically significant improvement over other methods (except FlowQA) with $p < 0.05$ tested by the Student's paired t-test. We can only do significance test on F1. 
}
\label{tab:results}
\vspace{-0.4cm}
\footnotesize
\begin{tabular}{@{}lllll@{}}
\toprule
Models                  & F1   & HEQ-Q & HEQ-D & Train Time (h) \\ \midrule
BiDAF++                 & 51.8 / 50.2  & 45.3 / 43.3  & 2.0 / 2.2 & -  \\
BiDAF++ w/ 2-Context    & 60.6 / 60.1  & 55.7 / 54.8  & 5.3 / 4.0 & -  \\
FlowQA                  & \textbf{64.6} / \textbf{64.1}  & \hspace{0.15cm}--\hspace{0.15cm}   / \textbf{59.6}  & \hspace{0.1cm}--\hspace{0.1cm} / \textbf{5.8} & 56.8 \\ \hline
BERT                    & 54.4 / \hspace{0.15cm}--\hspace{0.15cm}     & 48.9 / \hspace{0.15cm}--\hspace{0.15cm}     & 2.9 /\hspace{0.15cm}--\hspace{0.15cm} & 6.8     \\
BERT + PHQA             & 62.0 / \hspace{0.15cm}--\hspace{0.15cm}     & 57.5 / \hspace{0.15cm}--\hspace{0.15cm}     & 5.4 /\hspace{0.15cm}--\hspace{0.15cm} & 7.9     \\
BERT + PHA              & 61.8 / \hspace{0.15cm}--\hspace{0.15cm}     & 57.5 / \hspace{0.15cm}--\hspace{0.15cm}     & 4.7 /\hspace{0.15cm}--\hspace{0.15cm} & 7.2     \\
\textbf{BERT + HAE}              & \textbf{63.1}$^\ddagger$ / \textbf{62.4}  & \textbf{58.6} / \textbf{57.8}  & \textbf{6.0} / \textbf{5.1} & 10.1  \\ \bottomrule
\end{tabular}
\end{table}

We summarize our observations as follows. (1) Incorporating conversation history significantly boosts the performance in ConvQA. This is true for both BiDAF++ and BERT-based models. This not only suggests the importance of conversation history, but also shows the effectiveness of our history modeling approaches. (2) Our BERT-based ConvQA model outperforms BiDAF++. Furthermore, BERT with any of the history modeling methods outperform BiDAF++ w/ 2-Context. This shows the advantage of using BERT for ConvQA. (3) Prepending history turns with PHQA and PHA are both effective. The fact that they achieve similar performance suggests that history questions contribute little to the performance. This verifies our observation of the data that most follow-up questions are relevant to history answers. (4) Our HAE approach achieves better performance than simply prepending history turns. This indicates that HAE is more effective in modeling conversation history. (5) HAE manages to perform reasonably well with a relatively simple history modeling approach compared with the state-of-the-art method FlowQA.
(6) In addition to the model performance, we also compare the training efficiency. 
We observe our models are at least 5 times faster than FlowQA in training.\footnote{We use the code at \url{https://github.com/momohuang/FlowQA}, which was released by original authors. We set the batch size to 1 \textit{dialog} per batch to avoid memory issues.} Prepending history has little impact on training efficiency. Compared to PHQA, PHA is slightly faster because it only considers history answers. HAE is slightly slower than PH(Q)A but achieves considerable improvements. Compared to FlowQA, our HAE method achieves comparable performance with much higher training efficiency. 

\subsection{Impact of Conversation History}
\label{subsec:impact-of-history}
We then give an in-depth analysis on the impact of different amounts of conversation history with different history modeling approaches.


\begin{figure}[ht]
	\centering
	\begin{subfigure}[b]{0.24\textwidth}
        \includegraphics[width=\textwidth]{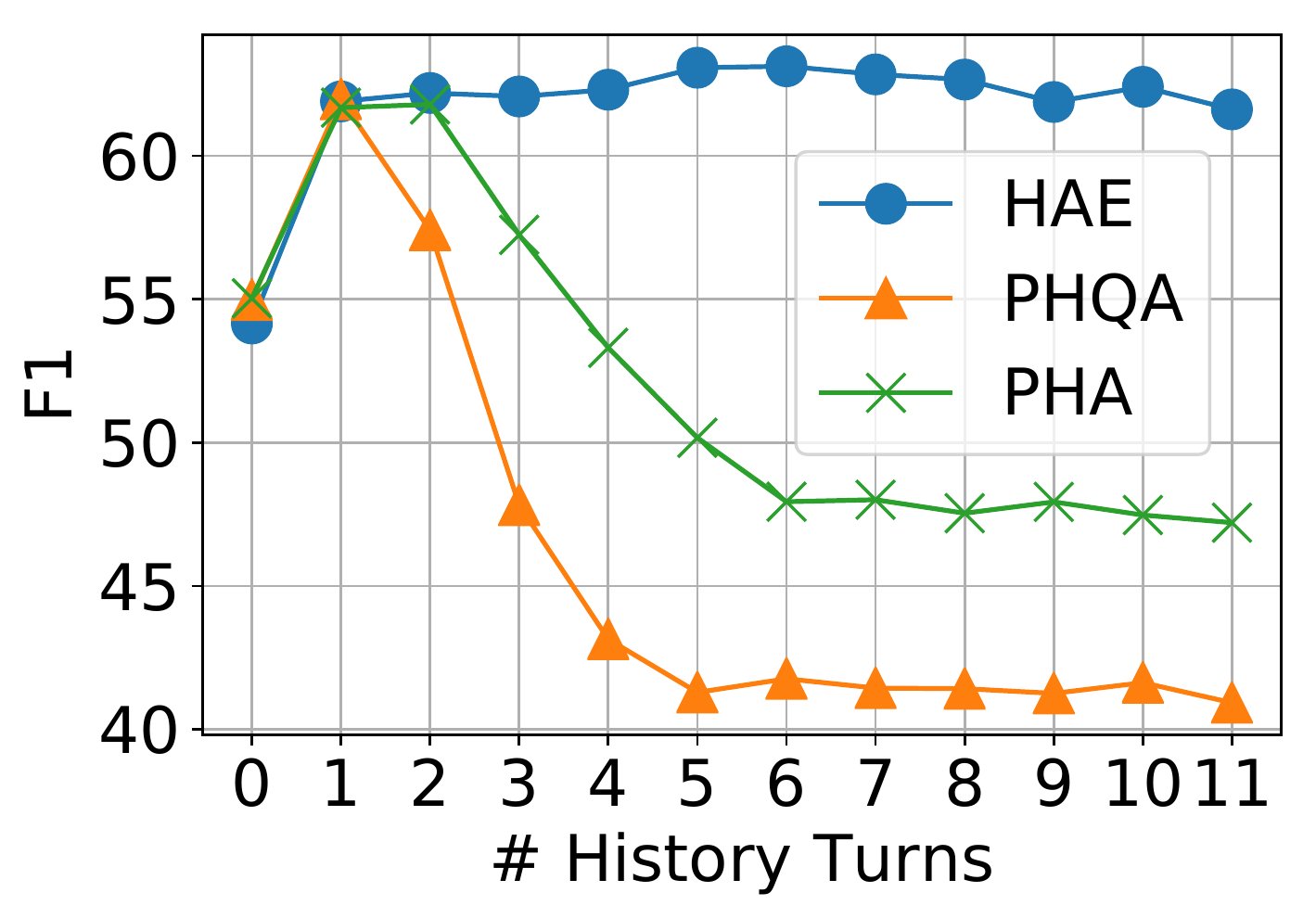}
		\vspace{-0.65cm}
		\label{fig:analyze-history-f1}
		\caption{F1}
	\end{subfigure}
	\begin{subfigure}[b]{0.23\textwidth}
        \includegraphics[width=\textwidth]{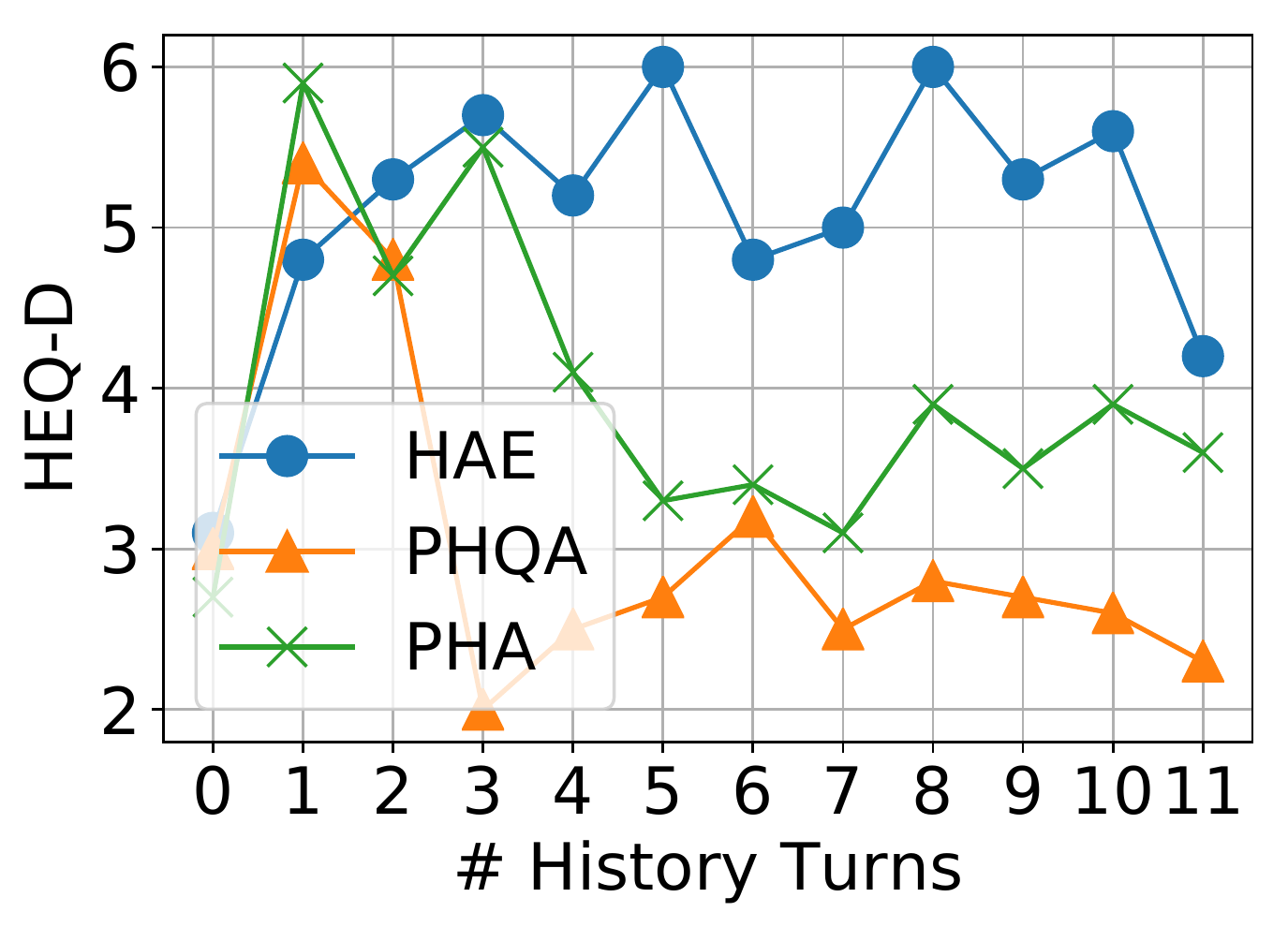}
        \vspace{-0.65cm}
		\label{fig:analyze-history-heqd}
		\caption{HEQ-D}
	\end{subfigure}
	\vspace{-0.85cm}
	\caption{Impact of different amounts of conversation history with different history modeling methods with BERT.}
	\label{fig:analyze-history}
\end{figure}

As presented in Figure~\ref{fig:analyze-history}, the most important observation is that our history answer embedding method can handle more conversation histories than simply prepending history turns, which shows the robustness of HAE. More importantly, the ability to model more history turns indeed gives some gains. This is based on the fact that HAE with 5 or 6 history turns gives the best performance. In addition, \citet{quac} also show that their answer marking method in BiDAF++ saturates at two turns. This further verifies the capability of our method to handle long conversation history.

Another interesting observation is that both PHQA and PHA show dramatic degradation as the number of history turns grows.\footnote{We used the same \textit{max question length} for all our BERT-based models as reported in Section~\ref{subsubsec:details}. We later discovered this hyper-parameter setting might be too strict for our BERT+PH(Q)A models to do a proper concatenation of history in some cases. When we relaxed this constraint, we observed that the degradation of BERT-PH(Q)A is relieved but still exists. PH(Q)A and HAE are both capable of incorporating a \textit{short} history. However, given the limited max sequence length for BERT, HAE is a more elegant and length-efficient method to incorporate a \textit{long} history.}
The best performance of PH(Q)A occurs at considering only 1 or 2 history turns. This is consistent with the results by \citet{coqa}, who use the same history modeling approach on Seq2Seq and DrQA~\cite{drqa}. The low performance when prepending a large amount of history suggests that a BERT-based ConvQA model is especially vulnerable to long prepended questions. This can be explained by the mechanism of constructing the input sequence as explained in Section~\ref{subsubsec:machine-comprehension}. A long prepended question shrinks the passage part in the input sequence and affect the answer prediction performance. This also explains the observation that PHQA drops faster as it also prepends history questions in addition to answers. These results show that history answer embedding is a better history modeling approach in a BERT-based ConvQA model. This is reasonable as HAE can be seamlessly integrated into BERT as shown in Figure~\ref{fig:cqa-model}.

%% file: 5_conclusions.tex
\section{Conclusions and Future work} 
\label{sec:conclusion}

In this work, we introduce a general framework for ConvQA to illustrate the two aspects of handling conversation history. We then propose a history answer embedding method to model conversation history in ConvQA. Extensive experiments show the effectiveness of our method. Finally, we perform an in-depth analysis to show the impact of different amounts of conversation history under different settings. Future work will consider to integrate our history modeling method with a learned history selection strategy for ConvQA.


%% file: 0_CQA.bbl
\begin{thebibliography}{20}
\providecommand{\natexlab}[1]{#1}
\providecommand{\url}[1]{\texttt{#1}}
\expandafter\ifx\csname urlstyle\endcsname\relax
  \providecommand{\doi}[1]{doi: #1}\else
  \providecommand{\doi}{doi: \begingroup \urlstyle{rm}\Url}\fi

\bibitem[Belkin et~al.(1994)Belkin, Cool, S., and Thiel]{Belkin1994CasesS}
N.~J. Belkin, C.~Cool, A.~S., and U.~Thiel.
\newblock {Cases , Scripts , and Information-Seeking Strategies : On the Design
  of Interactive Information Retrieval Systems}.
\newblock 1994.

\bibitem[Chen et~al.(2017)Chen, Fisch, Weston, and Bordes]{drqa}
D.~Chen, A.~Fisch, J.~Weston, and A.~Bordes.
\newblock {Reading Wikipedia to Answer Open-Domain Questions}.
\newblock In \emph{ACL}, 2017.

\bibitem[Choi et~al.(2018)Choi, He, Iyyer, Yatskar, Yih, Choi, Liang, and
  Zettlemoyer]{quac}
E.~Choi, H.~He, M.~Iyyer, M.~Yatskar, W.~Yih, Y.~Choi, P.~Liang, and L.~S.
  Zettlemoyer.
\newblock {QuAC: Question Answering in Context}.
\newblock In \emph{EMNLP}, 2018.

\bibitem[Croft and Thompson(1987)]{i3r}
W.~B. Croft and R.~H. Thompson.
\newblock {I3R: A new approach to the design of document retrieval systems}.
\newblock \emph{JASIS}, 38:\penalty0 389--404, 1987.

\bibitem[Devlin et~al.(2018)Devlin, Chang, Lee, and Toutanova]{bert}
J.~Devlin, M.-W. Chang, K.~Lee, and K.~Toutanova.
\newblock {BERT: Pre-training of Deep Bidirectional Transformers for Language
  Understanding}.
\newblock \emph{CoRR}, 2018.

\bibitem[Gao et~al.(2018)Gao, Galley, and Li]{Gao2018NeuralAT}
J.~Gao, M.~Galley, and L.~Li.
\newblock {Neural Approaches to Conversational AI}.
\newblock In \emph{SIGIR}, 2018.

\bibitem[Huang et~al.(2018)Huang, Choi, and Yih]{flowqa}
H.-Y. Huang, E.~Choi, and W.~Yih.
\newblock {FlowQA: Grasping Flow in History for Conversational Machine
  Comprehension}.
\newblock \emph{CoRR}, 2018.

\bibitem[Kotov and Zhai(2010)]{Kotov2010TowardsNQ}
Alexander Kotov and ChengXiang Zhai.
\newblock Towards natural question guided search.
\newblock In \emph{WWW}, 2010.

\bibitem[Nguyen et~al.(2016)Nguyen, Rosenberg, Song, Gao, Tiwary, Majumder, and
  Deng]{Marco}
T.~Nguyen, M.~Rosenberg, X.~Song, J.~Gao, S.~Tiwary, R.~Majumder, and L.~Deng.
\newblock {MS MARCO: A Human Generated MAchine Reading COmprehension Dataset}.
\newblock \emph{CoRR}, abs/1611.09268, 2016.

\bibitem[Qu et~al.(2018)Qu, Yang, Croft, Trippas, Zhang, and
  Qiu]{Qu2018AnalyzingAC}
C.~Qu, L.~Yang, W.~B. Croft, J.~R. Trippas, Y.~Zhang, and M.~Qiu.
\newblock {Analyzing and Characterizing User Intent in Information-seeking
  Conversations}.
\newblock In \emph{SIGIR}, 2018.

\bibitem[Qu et~al.(2019)Qu, Yang, Croft, Zhang, Trippas, and
  Qiu]{UserIntentPred}
C.~Qu, L.~Yang, W.~B. Croft, Y.~Zhang, J.~R. Trippas, and M.~Qiu.
\newblock {User Intent Prediction in Information-seeking Conversations}.
\newblock \emph{CoRR}, 2019.

\bibitem[Rajpurkar et~al.(2016)Rajpurkar, Zhang, Lopyrev, and Liang]{squad}
P.~Rajpurkar, J.~Zhang, K.~Lopyrev, and P.~Liang.
\newblock {SQuAD: 100, 000+ Questions for Machine Comprehension of Text}.
\newblock In \emph{EMNLP}, 2016.

\bibitem[Reddy et~al.(2018)Reddy, Chen, and Manning]{coqa}
S.~Reddy, D.~Chen, and C.~D. Manning.
\newblock {CoQA: A Conversational Question Answering Challenge}.
\newblock \emph{CoRR}, abs/1808.07042, 2018.

\bibitem[Seo et~al.(2016)Seo, Kembhavi, Farhadi, and Hajishirzi]{bidaf}
M.~J. Seo, A.~Kembhavi, A.~Farhadi, and H.~Hajishirzi.
\newblock {Bidirectional Attention Flow for Machine Comprehension}.
\newblock \emph{CoRR}, abs/1611.01603, 2016.

\bibitem[Trippas et~al.(2018)Trippas, Spina, Cavedon, Joho, and
  Sanderson]{Trippas2018InformingTD}
J.~R. Trippas, D.~Spina, L.~Cavedon, H.~Joho, and M.~Sanderson.
\newblock {Informing the Design of Spoken Conversational Search: Perspective
  Paper}.
\newblock In \emph{CHIIR}, 2018.

\bibitem[Vaswani et~al.(2017)Vaswani, Shazeer, Parmar, Uszkoreit, Jones, Gomez,
  Kaiser, and Polosukhin]{transformer}
A.~Vaswani, N.~Shazeer, N.~Parmar, J.~Uszkoreit, L.~Jones, A.~N. Gomez,
  L.~Kaiser, and I.~Polosukhin.
\newblock {Attention Is All You Need}.
\newblock In \emph{NIPS}, 2017.

\bibitem[Yang et~al.(2017)Yang, Zamani, Zhang, Guo, and
  Croft]{Yang2017NeuralMM}
L.~Yang, H.~Zamani, Y.~Zhang, J.~Guo, and W.~B. Croft.
\newblock {Neural Matching Models for Question Retrieval and Next Question
  Prediction in Conversation}.
\newblock \emph{CoRR}, 2017.

\bibitem[Yang et~al.(2018)Yang, Qiu, Qu, Guo, Zhang, Croft, Huang, and
  Chen]{Yang2018ResponseRW}
L.~Yang, M.~Qiu, C.~Qu, J.~Guo, Y.~Zhang, W.~B. Croft, J.~Huang, and H.~Chen.
\newblock {Response Ranking with Deep Matching Networks and External Knowledge
  in Information-seeking Conversation Systems}.
\newblock In \emph{SIGIR}, 2018.

\bibitem[Zhang et~al.(2018)Zhang, Chen, Ai, Yang, and
  Croft]{Zhang2018TowardsCS}
Y.~Zhang, X.~Chen, Q.~Ai, L.~Yang, and W.~B. Croft.
\newblock {Towards Conversational Search and Recommendation: System Ask, User
  Respond}.
\newblock In \emph{CIKM}, 2018.

\bibitem[Zhu et~al.(2018)Zhu, Zeng, and Huang]{sdnet}
C.~Zhu, M.~Zeng, and X.~Huang.
\newblock {SDNet: Contextualized Attention-based Deep Network for
  Conversational Question Answering}.
\newblock \emph{CoRR}, 2018.

\end{thebibliography}
